# DEVELOPMENT OF THE LOW ENERGY ACCELERATOR FOR KOMAC

B.H. Choi, J.M. Han, Y.S. Cho, K.S. Kim, Y.J. Kim, KAERI, Taejon, Korea
I.S. Ko, Postech, Pohang, Korea
K.S Chung Hanyang Univ, Seoul, Korea
K.H. Chung, Seoul Nat. Univ., Seoul, Korea

*Abstract*

KAERI (Korea Atomic Energy Research Institute) has been performing the project named KOMAC (Korea Multi-purpose Accelerator Complex) within the frame work of national mid and long term nuclear research plan. The final objectives of KOMAC is to build a 20-MW (1 GeV and 20 mA) cw (100% duty factor) proton linear accelerator. As the first stage, the low energy accelerator up to 20 MeV is being developed in KTF (KOMAC Test Facility). The low energy accelerator consists of an injector, RFQ, CCDTL, and RF systems. The proton injector with Duoplasmatron ion source has been developed, and the LEBT with solenoid lens is under development. The RFQ linac that will accelerate a 20mA proton beam from 50keV to 3MeV has been designed and is being fabricated. The RF system for RFQ is being developed, and the CCDTL up to 20MeV is being designed. The status of the low energy accelerator will be presented.

## 1 INTRODUCTION

The KOMAC accelerator has been designed to accelerate a 20 mA cw proton/H- with the final energy 1GeV cw super-conducting linac [1]. In the first stage of the project, we are developing cw accelerating structure up to 20MeV, and operate the accelerator in 10% duty pulse mode. After the first stage, we will challenge the cw operation of the accelerator. The 20MeV proton accelerator is constructing in the KTF (KOMAC Test Facility), and will be commissioned in 2003. After the commissioning, KTF will provide the proton beam for the many industrial applications.

In the KTF, we are developing the proton injector, LEBT, 3MeV RFQ, 20MeV CCDTL, and RF system. The proton injector is already developed, and the 3MeV RFQ will be constructed in this fiscal year. Also we have a plan to develop the basic Super-Conducting cavity technology in the KTF for the second stage super-conducting accelerator of the KOMAC. Fig. 1 shows the plan of the KTF and Fig. 2 shows the status of the KTF.

The status of the low energy accelerator developments in KTF will be introduced in this paper.

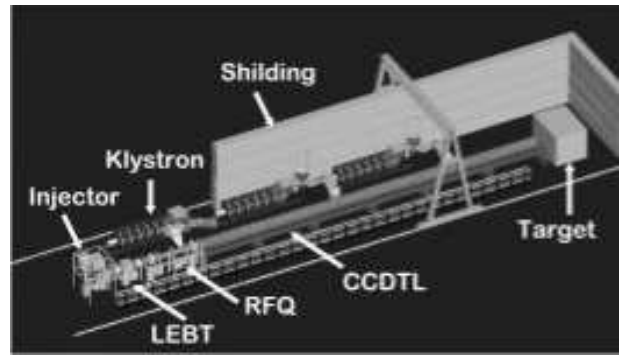

Figure 1: Plan of KTF 20MeV Accelerator

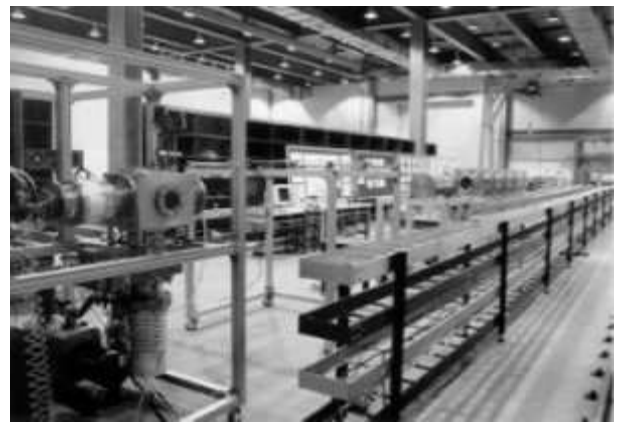

Figure 2: Status of KTF 20MeV Accelerator

## 2 PROTON INJECTOR [2]

For 20 mA proton beam at the final stage, KOMAC requires the ion source with the proton beam current of 30 mA at the extraction voltage of 50 kV. Normalized rms emittance of less than 0.3 $\pi$ mm·mrad is also required for good matching of ion beam into RFQ.

The proton injector with a duoplasmatron ion source is shown in Fig. 2 (left side). The system is composed of an accelerating high voltage power supply, ion source power supplies in a high voltage deck, gas feeding system, and vacuum system.

The injector has reached beam currents of up to 50 mA at 50 kV extraction voltage with 150 V, 10 A arc power. The extracted beam has a normalized emittance of 0.2 $\pi$ mm·mrad from 90 % beam current and proton fraction of over 80 %. The proton fraction is measured with deflection magnet and scanning wire.

The beam can be extracted without any fluctuation in beam current, with a high voltage arcing in 4 hours. The cathode lifetime is about 40hr. To increase the filament lifetime, it is necessary to lower the arc current or to change the tungsten filament to other cathode such as oxide cathode.

## 3 LEBT

Low-energy beam transport (LEBT) consists of two solenoids, two steering magnets, diagnostic system, beam control system, and funnelling system to transports and matches the $H^+$ for 20mA and $H^-$ for 3mA, beams from the ion source into the RFQ. The main goal of the LEBT design is to minimise beam losses. The design codes used are TRACE 3D and PARMTEQM. The PARMTEQM-simulated solenoid settings are B=2800G and B=3900G, the RFQ transmission rate is more than 90%. Two solenoid magnets constructed are 20.7cm-long, 16cm-i.d., are surrounded by a low carbon steel and provide dc fields ≤5000G on the axis. During the winter of 2000, we will test the LEBT to obtain a proper matching condition with the RFQ.

## 4 RFQ [3]

The KTF RFQ bunches, focuses, and accelerates the 50keV $H^+/H^-$ beams, and derives a 3.0MeV beam at its exit, bunched with a 350MHz. The RFQ is a 324cm-long, 4-vanes type, and consists of 56 tuners, 16 vacuum ports, 1 coupling plate, 4 rf drive couplers, 96 cooling passages, and 8 stabiliser rods. The RFQ is machined of OFH-Copper, integrate from separate four sections which are constructed by using vacuum furnace brazing. The RF system for the RFQ is operated with 350MHz at 100% duty-factor by one klystron of 1MW.

Its design was completed. In the RFQ design, a main issue is to accelerate the mixed $H^+/H^-$ beam at the same time. The motion of the mixed $H^+/H^-$ beam into the RFQ has been studied by using a time marching beam dynamics code QLASSI. The longitudinal beam loss increases with the concentration of negative ions by the bunching process which is distributed by attractive forces when the ratio of $H^-$ is more than 30%.. Because of the space charge compensation in the low energy sections, the transverse beam loss decreases with the mixing ratio of $H^-$.

The average RFQ cavity structure power by rf thermal loads is 0.35 MW and the peak surface heat flux on the cavity wall is 0.13 $MW/m^2$ at the high energy end. In order to remove this heat, we consider 24 longitudinal coolant passages in each of the sections. In the design of the coolant passages, we considered the thermal behaviour of the vane during CW operation and manufacturing costs. The thermal and structure analysis is studied with SUPERFISH and ANSYS codes. The temperature of the coolant passages on the cavity wall is varied to maintain the cavity on resonance frequency.

As a test bed for 3MeV RFQ, the design, construction, electrical test, and vacuum test of the 0.45MeV RFQ have been finished. Design of the RFQ was done by KAERI and POSTECH, a fabrication was done at Dae-Ung Engineering Company and VITZRO TECH Co., Ltd. A difficult process in the fabrication of the RFQ was to braze. Because of the leak of the brazing surface and the strain of the RFQ structure by the furnace heat, it is important to determine an appropriate shape of the brazing area. To determine it, we have performed two brazing test. Fig. 3 shows a 96.4cm long 0.45MeV RFQ which was brazed in a vacuum furnace. The RFQ was brazed in a vertical orientation with LUCAS BVag-8, AgCu alloy with a liquid temperature of 780 $^\circ$C. Testing of the brazed RFQ showed it to be leak-tight.

The coolant passages in the cavity wall and vane area were drilled with a deep hole. The entrances of deep holes at the vane end was brazed. The exact dimension of the undercut was determined empirically by cutting a vane of the hot model which was fabricated of the OFHC. In the case of the RFQ with a modulated vane tip, the resonant frequency of the RFQ cavity linearly decreases with undercut depth. However, in the case of the RFQ cold model with a constant vane tip, the resonant frequency of the RFQ cavity non-linearly decreases with undercut depth.

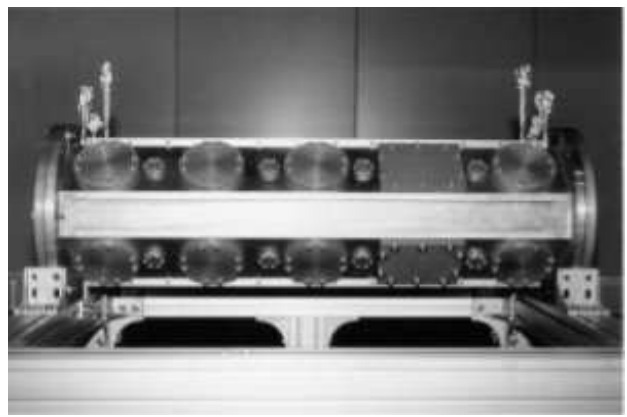

Figure 3: A brazed 0.45MeV RFQ

The 3MeV RFQ cold model of aluminium was fabricated and tested. A low-level RF control system, which maintains proper amplitude(within ±1%) and phase(within ±1°), has been designed. A cold model of a tuner has been fabricated and is being tested.

Assembly works of the 3MeV RFQ will be done in March, 2001.

## 5 CCDTL [4]

CCDTL will accelerate the 3MeV 20mA proton beam to the energy of 20MeV. The structure design of CCDTL is based on the 100% duty factor.

Table 1: Major Design Parameters of CCDTL cavity

- Structure : 700MHz CCDTL
- Length : 25m
- Aperture Diameter : 10/15mm
- No of EMG : 130 (8 βλ FODO)
- Total Structure Power : 1.15MW
- Sturcture Power per length : 50kW/m avg.
- Surface E : <0.9 Kilpatrick

The CCDTL cold models are fabricated to check the design, the tuning method, and the coupling coefficients and the fabrication method. The measured resonant frequency is 700.8 MHz without air and humidity compensation. The measured Q value of the cavity without brazing is 87% of the calculated Q by SUPERFISH without any surface cleaning. The super-drilled coolant path is well fabricated, and this type cooling method will be used for the CCDTL construction.

The field profile is measured with a bead perturbation method. The field measurement system is shown in Fig. 4. A 2mm diameter and 2mm long alumina cylinder is used for the bead. The stepping motor drive system controls the position of the bead with an accuracy of 0.2mm. The frequency shift is measured with a network analyzer (HP4306A/85064A). Because the temperature controlled room is not available, the measurement was carried with the careful check of the unperturbed resonance frequency before and after the experiment.

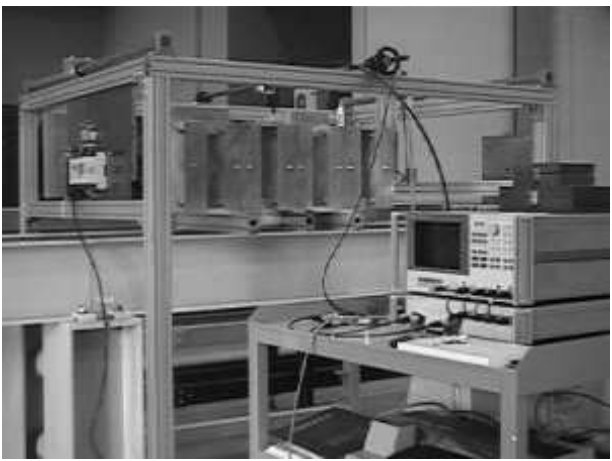

Figure 4: CCDTL Field Measurement

Fig. 5 shows the one measured field profile in one cavity of the aluminium cold model. The measured field profile in a cavity agrees with the calculated profile. But, the field uniformity in the multi-cavity is not good. It is necessary to increase the field uniformity by the fine tuning of the cavity. This will be done with the brazed copper cold model that will be fabricated in this year. The copper model will be fabricated with the study. As a back-up of the CCDTL, the design study for conventional DTL will be performed.

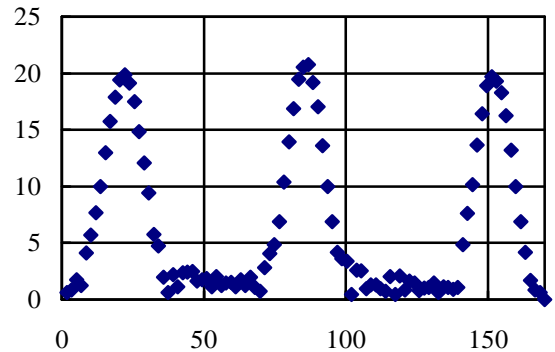

Figure 5: Measured Field Profile in One Cavity
(x: Position(mm), y: Field(Arb.))

## 6 SUMMARY

The low energy proton accelerator for KTF is designed. The proton injector can provide the proton beam for RFQ. The RFQ is fabricating and will be tested with 1MW RF system. The CCDTL is studied with cold models, and the hot model will be fabricated.

### ACKNOWLEDGEMENT

This work was supported by the Korea Ministry of Science and Technology.